\documentclass[]{aastex631}
\pdfoutput=1 
\usepackage{amsmath,amstext,amssymb,mathtools,graphicx,color}

\usepackage{graphicx}	
\usepackage{amsmath}	
\usepackage{amssymb}	
\usepackage{xcolor}
\usepackage{lipsum}
\usepackage[normalem]{ulem}

\submitjournal{ApJ}

\shorttitle{Supernova Mixing Diagnostics}
\shortauthors{Fryer et al.}

\graphicspath{{./}{figures/}}

\begin{document}

\title{Multi-Messenger Diagnostics of the Engine behind Core-Collapse Supernovae}

\correspondingauthor{Chris Fryer}
\email{fryer@lanl.gov}

\author[0000-0003-2624-0056]{Christopher~L. Fryer}
\affiliation{Center for Theoretical Astrophysics, Los Alamos National Laboratory, Los Alamos, NM, 87545, USA}
\affiliation{The University of Arizona, Tucson, AZ 85721, USA}
\affiliation{Department of Physics and Astronomy, The University of New Mexico, Albuquerque, NM 87131, USA}
\affiliation{The George Washington University, Washington, DC 20052, USA}

\author[0000-0002-2942-3379]{Eric Burns}
\affiliation{Department of Physics \& Astronomy, Louisiana State University, Baton Rouge, LA 70803, USA}

\author[0000-0000-0000-0000]{Aimee Hungerford}
\affiliation{Center for Theoretical Astrophysics, Los Alamos National Laboratory, Los Alamos, NM, 87545, USA}

\author[0000-0000-0000-0000]{Samar Safi-Harb}
\affiliation{Department of Physics and Astronomy, University of Manitoba, Winnipeg, MB R3T 2N2, Canada}

\author[0000-0003-3265-4079]{R.~T. Wollaeger}
\affiliation{Center for Theoretical Astrophysics, Los Alamos National Laboratory, Los Alamos, NM, 87545, USA}

\author[0000-0003-4496-7128]{Richard~S. Miller}
\affiliation{Johns Hopkins University Applied Physics Laboratory, Laurel, MD 20723, USA}

\author[0000-0002-6548-5622]{Michela Negro}
\affiliation{University of Maryland, Baltimore County, Baltimore, MD 21250, USA}
\affiliation{NASA Goddard Space Flight Center, Greenbelt, MD 20771, USA}
\affiliation{Center for Research and Exploration in Space Science and Technology, NASA/GSFC, Greenbelt, MD 20771, USA}

\author[0000-0002-5490-2689]{Samalka Anandagoda}
\affiliation{Center for Theoretical Astrophysics, Los Alamos National Laboratory, Los Alamos, NM, 87545, USA}
\affiliation{Clemson University, Department of Physics \& Astronomy, Clemson, SC 29634-0978, USA}

\author[0000-0002-8028-0991]{Dieter H. Hartmann}
\affiliation{Clemson University, Department of Physics \& Astronomy, Clemson, SC 29634-0978, USA}



\begin{abstract}

Core-collapse supernova explosions play a wide role in astrophysics by producing compact remnants (neutron stars, black holes) and the synthesis and injection of many heavy elements into their host Galaxy.  Because they are produced in some of the most extreme conditions in the universe, they can also probe physics in extreme conditions (matter at nuclear densities and extreme temperatures and magnetic fields).  To quantify the impact of supernovae on both fundamental physics and our understanding of the Universe, we must leverage a broad set of observables of this engine.  In this paper, we study a subset of these probes using a suite of 1-dimensional, parameterized mixing models: ejecta remnants from supernovae, ultraviolet, optical and infra-red lightcurves, and transient gamma-ray emission.  We review the other diagnostics and show how the different probes tie together to provide a more clear picture of the supernova engine\footnote{Join us in improving and evolving this document through active community engagement. \\Instructions are provided at this link: \url{https://github.com/clfryer/MM-SNe}.}.

\end{abstract}


\section{Introduction} 
\label{sec:intro}

Supernova (SN) 1987A marked a major inflection point in our understanding of core-collapse supernovae.  Observations of the progenitor and neutrinos proved that explosions could indeed occur from the collapse of the core of a massive star down to a compact remnant (neutron star or black hole), confirming the basic core-collapse supernova engine.  But it also brought a number of surprises.  One surprise was that material believed to be produced in the innermost ejecta was somehow mixed into the outer layers of the star.  For example, $^{56}$Ni is primarily produced near the proto-neutron star and is believed to be in the innermost ejecta.  Gamma rays from the decay of this $^{56}$Ni are initially trapped in the flow and only when the ejecta expand to low densities can these gamma rays escape and be observed.  The gamma rays from SN 1987A began to escape (and were observed) much earlier than expected from spherically-symmetric models~\citep{1988ApJ...329..820P}.  In addition, this material is expected to be slow moving (producing narrow lines), but observations of the iron lines (decay product of $^{56}$Ni) were much broader than expected~\citep{1990MNRAS.242..669S,1990ApJ...360..257H}.  

Based on this observational evidence, \cite{herant94} developed a model where convection above the proto-neutron star would both enhance the conversion of internal energy produced in the bounce of the stellar core when it reaches nuclear densities and reduce the pressure of the infalling material.  Their first 2-dimensional models demonstrated that convection above the proto-neutron star could help drive a supernova explosion.  This convection provided a natural explanation for supernova energetics set by the ram pressure of the infalling star~\citep{2012ApJ...749...91F} while its asymmetries explained not only the features of SN 1987A, but also the properties of supernova ejecta remnants~\citep{2014Natur.506..339G} and properties (i.e. kick velocities) of the compact remnants~\citep{1995PhR...256..117H,2013A&A...552A.126W,2022ApJ...926....9J}. Since these results, an increasing number of simulations have produced explosions under this paradigm, and it has slowly gained traction to become the leading theory model for the supernova engine~\citep{2007ApJ...659.1438F,2014ApJ...786...83T,2015ApJ...807L..31L,2015ApJ...808L..42M,2018SSRv..214...33B,2021ApJ...921...28F}.   The combined observational success and theoretical confirmation has caused this convection to become part of the standard paradigm in supernovae.  But the details of this engine remain a matter of intense debate and further observations are needed to help truly understand this engine~\citep[for a review, see][]{2022ApJ...931...94F}.

Other engine proposals exist, many invoking the collapse of rapidly-rotating stellar cores that either produce disks around the newly formed neutron stars or produce a rapidly-rotating neutron star that, if magnetized, can produce a strong outflow~\cite[for a review, see][]{2022IJMPD..3130005F,2022arXiv220804875S}.  Although the standard paradigm suggests these engines are only a small subset of the observed supernovae but the role of this engine remains a matter of great debate in astrophysics.

To date, the strongest observational constraints on the asymmetries in the supernova engine has been from observations of supernova remnants.  These observations measure the distribution of elements produced in the innermost ejecta, either from the material in the convective region or in material just above this convective region that is heated by the supernova shock as it is launched.  This ejecta then cools as it explosively expands producing the elements observed in the remnants~\citep{2014Natur.506..339G,2017ApJ...834...19G}.  Observed through the hard X-ray/$\gamma$-rays emitted in nuclear decay, these remnant distributions provide a clean probe of the engine asymmetries.  Both future morphology measurements and more detailed nucleosynthetic yield estimates have the potential to further constrain the nature of the convection engine. 

In this paper, we outline the broad range of diagnostics that can be used to probe the supernova engine (Section~\ref{sec:diag}).  We provide a review of those diagnostics that are beyond the scope of this paper and focus on a subset of these diagnostics using a grid of supernova models (described in Section~\ref{sec:grid}).  In Section~\ref{sec:SNR}, we leverage our grid of 1-dimensional models (as well as some 3-dimensional models) to determine how observations of supernova remnants probe the supernova engine and the outward mixing of material.  We also apply this grid to study the ability for supernova light-curves (Section~\ref{sec:lc}) and prompt gamma-ray emission (Section~\ref{sec:gray}) to probe the supernova engine.  We summarize the multi-diagnostic constraints in Section~\ref{sec:corr}.

\section{Diagnostics of the Supernova Engine}
\label{sec:diag}

\begin{figure}
    \centering
    \includegraphics[width=7.0in]{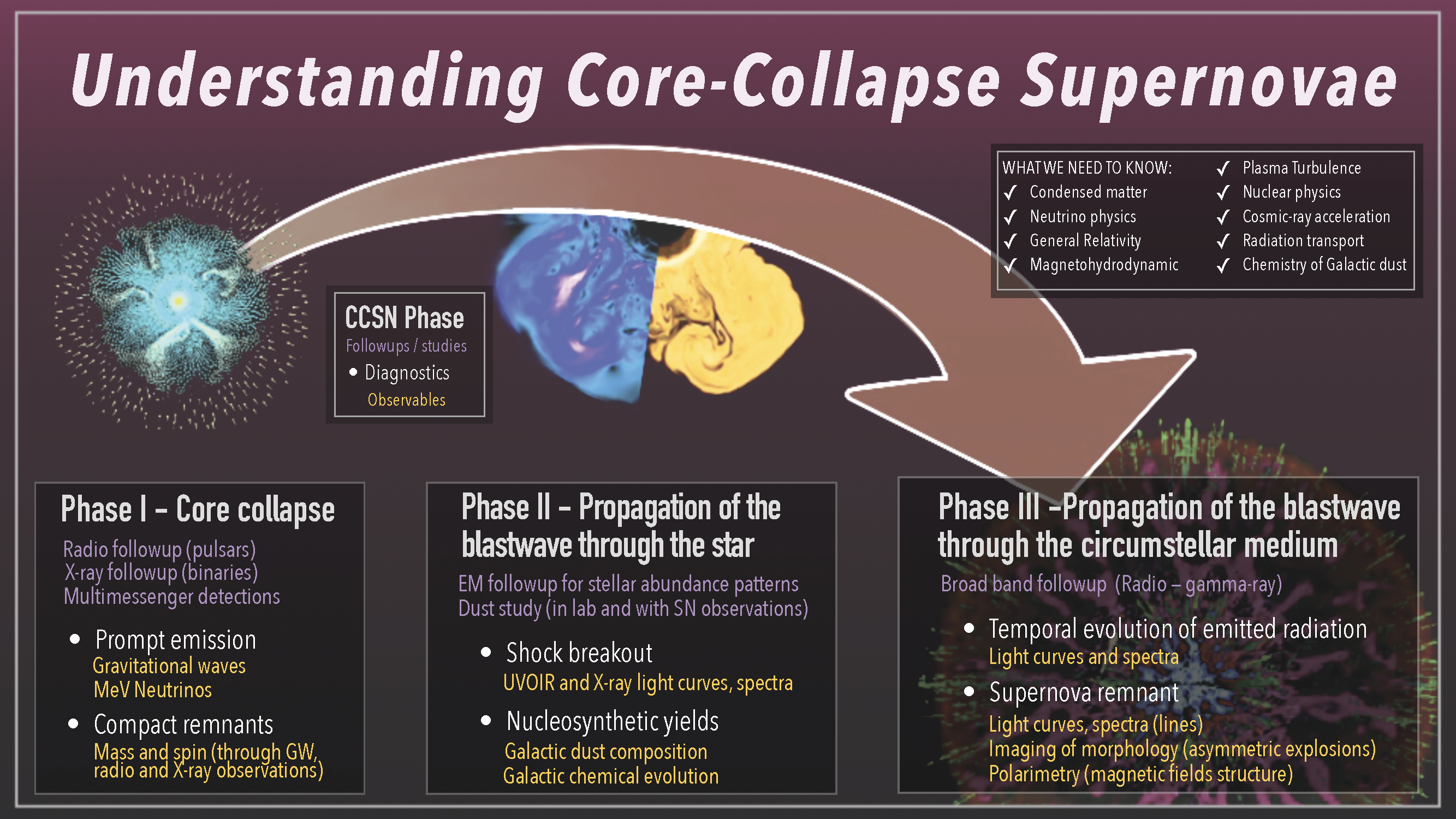}
    \caption{The broad set of observations used to probe the supernova engine.  To do these studies with a full understanding of the uncertainties requires combining multiple studies using a variety of codes implementing different physics in different regimes.  Gravitational waves and neutrino emission peak near, or shortly after, core bounce ($t_0$).  Shock breakout occurs $t_0$+30-300\,s (Type Ib/c supernovae) or $t_0$+$\sim 10,000$\,s (Type IIP supernovae).  Depending on the amount of asymmetry-driven mixing, MeV Gamma-Ray can rise shortly thereafter (peaking anywhere between 5\,d to beyond 100\,d.)  In UVOIR bands, shock interactions and shock cooling dominates after shock breakout and peaks at roughly 10-20\,d.}
    \label{fig:diag}
\end{figure}

Connecting the supernova engine to potential diagnostics requires coupling together a broad range of physics expertise and numerical methods.  Figure~\ref{fig:diag} outlines the different phases of supernovae from progenitor to remnants and the diagnostics that can probe characteristics of the supernova engine.  In this paper, we present new studies of three diagnostics:  supernova light-curves, gamma-ray emission, and ejecta remnants.  However, in Section~\ref{sec:corr}, we discuss the correlations of these three diagnostics with the broad set of additional supernova diagnostics.  In this section, we review these additional diagnostics with a focus on their potential to constrain the nature of the supernova engine.  We divide these constraints into two categories:  direct probes that require a single layer of modeling to tie to the central engine (but are typically very rare) and indirect probes that  require multiple layers of modeling to tie to the central engine (but are typically more common). 

\subsection{Direct Probe:  Gravitational Waves}

Gravitational waves are the most direct probe of the supernova engine and can be used to both study the angular momentum in the collapsing stellar core and the development and magnitude of convection in both the proto-neutron star and in the convective engine region (above the proto-neutron star and below the infall shock).  Gravitational waves can also be produced through highly-asymmetric neutrino emission. \cite{2011LRR....14....1F} reviewed potential sources and processes driving gravitational waves from core-collapse and we will summarize this review here, focusing on the detection probabilities for the major sources.  For this discussion, we focus on the LIGO (O3), Cosmic Explorer and Einstein Telescope sensitives using the analysis of the summary paper:  \citet{2021arXiv210909882E}.  

Much of the confusion on the detectability of gravitational waves lies in the assumptions behind the angular momentum of the collapsing core.  This angular momentum is conserved in the collapse causing this material to spin up.  In the most extreme cases, the core can fragment or form bar instabilities as it collapses~\citep{2002ApJ...565..430F}.  In this extreme case, LIGO (operating at 'O3' sensitivities) could have detected the collapse/bounce gravitational wave signal out to 10\,Mpc and Cosmic Explorer will detect the gravitational wave signal out to 100\,Mpc~\citep{2021arXiv210909882E}.  The LIGO detector is already capable to place limits on the extreme rotations needed to produce bar instabilities.  However, the extreme rotations needed to produce these instabilities is well beyond what any stellar model (single or in a binary) has ever produced.

A more likely scenario lies in the fast-spinning stellar models often invoked to explain pulsars born with fast spin periods, magnetar-driven explosions, collapsar gamma-ray bursts and rapidly spinning black holes in X-ray binaries.  Based on observations of these phenomena, we know that these are rare~\citep{2019EPJA...55..132F}, but the fraction of stars with this rapid rotation could be as high as 10\%.   Based on the gravitational wave signals predicted by rotating model~\cite[for a review, see][]{10.1093/mnrasl/sly008,1997A&A...317..140M,2004ApJ...609..288F,2011LRR....14....1F}, LIGO (O3) could detect supernovae out to 1\,Mpc and Cosmic Explorer would detect the signals out to 10\,Mpc.  After multiple years, a Cosmic Explorer-class detector could firmly set the fraction of these rapidly-rotating collapses, placing limits on the role fast-rotating magnetars play in producing supernovae and gamma-ray bursts.

For slowly- or non-rotating systems, the gravitational wave signal instead probes the growth timescales and magnitude of the convection.  Although convection in the proto-neutron star can possibly be detected~\citep{2011LRR....14....1F}, the bulk of the current work focuses on the convective region between the proto-neutron star and the infalling shock~\citep{2021MNRAS.500..696N,2009A&A...496..475M}.  This convection region is what drives the explosion under the current supernova engine paradigm~\citep{herant94}.  This signal is much weaker than rotating collapses and is detectable with LIGO within a few kpc.  However, even with Cosmic Explorer, the detection of these slowly rotating signals beyond the local group will be difficult~\citep{PhysRevD.103.023005}.

In Summary, the gravitational wave signal can differentiate between rotational and convective engines, but this will require multiple years with next generation (e.g. Cosmic Explorer) graviational wave detectors.  For a Galactic supernova, gravitational waves will probe the growth rate and magnitude of the convection.

\subsection{Direct Probe:  Neutrinos}

The detection of neutrinos from SN 1987A~\citep{1987PhRvL..58.1494B,1987EL......3.1315A} proved definitively that at least some supernovae were produced from the collapse of the core of a massive star down to nuclear densities.  Coupled with observations of the progenitor of SN 1987A~\citep[e.g.][]{1987IAUC.4319....1W,1987IAUC.4333....1S,1987IAUC.4348....1W}, these neutrinos solidified the core-collapse supernova paradigm.  These neutrino observations have also been used to probe a broad range of physical properties of neutrinos including the neutrino mass, magnetic moments, neutrino mixing and oscillations~\citep{1987PhLB..196..267S,1988PhLB..200..115F,1988PhRvL..61...27B,1996PhRvD..54.1194J,2001PhLB..504..301M}.  

The detection of neutrinos from supernovae have focused on two different sources:  thermal emission from the collapse and bounce of the stellar core and non-thermal neutrino production during shock interactions with the circumstellar medium.  The latter is a potential probe of the conditions in the circumstellar medium and the physics behind the production of a wide range of high energy particles.  These high-energy ($>100\,{\rm GeV}$), non-thermal neutrinos can be detected by the IceCube facility in the south pole~\citep{2022icrc.confE1116N}.  If supernovae produce ultra-high energy neutrinos (neutrino energies above $10^{18}$\,eV) in these shocks, they may be detected by missions such as ANtarctic Impulsive Transient Antenna (ANITA)~\citep{2021JCAP...04..017A}. 

To probe the supernova engine itself, astronomers must focus on the thermal neutrinos emitted during collapse, explosion and proto-neutron star cooling phases.  Observations of the bounce and proto-neutron star cooling phases can be used to better understand the nature of the compact object, the potential formation of a black hole and the dense nuclear physics shaping that object~\citep[for reviews, see][]{2016NCimR..39....1M,2017hsn..book.1605R,2021PhRvD.103b3016L,2023arXiv230801403F}.  Neutrino observations can also probe open questions in neutrino physics including neutrino oscillations~\citep{2006PhRvD..74l3004D,2010ARNPS..60..569D}.  A suite of detectors are being developed and enhanced to detect electron, $\mu$ and $\tau$ neutrinos including JUNO, DUNE and Hyper-Kamiokande.  For a Galactic supernova 10\,kpc from us, we expect roughly 250 neutrino detections from Super-K~\citep{2021PhRvD.103b3016L}, allowing a detailed study of both the supernova engine and nuclear/neutrino physics.  Unfortunately, thermal neutrinos from supernovae will not be observed beyond roughly 100\,kpc~\citep{2021ApJ...916...15A}, limiting any supernova observations to the Milky Way and the local group.

Combining neutrinos and gravitational wave signatures can provide even further constraints~\citep{2017ApJ...851...62K,2020ApJ...898..139W} and we will discuss this further in Section~\ref{sec:corr}. 

\subsection{Indirect Probe:  Shock Breakout Observations}

The convective engine produces an asymmetric explosion and these explosion asymmetries in the engine propagate through the star and then into the circumstellar medium surrounding the star.  Throughout this evolution, it becomes increasingly difficult to disentangle the explosion asymmetries from new instabilities that develop as the shock propagates through the star and then the circumstellar medium.  As we will discuss in Section~\ref{sec:SNR}, these new instabilities can confuse the nature of the engine asymmetries.  The shock breakout diagnostic refers to observations as soon as the supernova blastwave breaks out of the star, prior to developing asymmetries from the propagation through the circumstellar medium.

The best-known example of shock breakout observations was the serendipitous detection of this emission in SN 2008D~\citep{soderberg2008extremely}.  Using simple spherically-symmetric models, scientists estimated the stellar radius.  However, improved models have shown that the analysis of such observations is much more difficult once explosion asymmetries~\citep{2021MNRAS.508.5766I}, stellar asymmetries~\citep{2022ApJ...933..164G} and wind asymmetries~\citep{2020ApJ...898..123F} are included.  These different effects lead to a broad set of emission energies from UV to X-ray.  The XMM-Newton satellite has discovered a number of potential shock breakout events~\citep{2020ApJ...896...39A}.  The {\it UltraSAT}~\citep{2022arXiv220800159B} and CASTOR~\citep{2017XRS....46..303M} satellites are designed to increase the number of shock breakout detections in the UV.  But to disentangle all of this physics, joint UV and X-ray observations are required.  The {\it SIBEX} mission~\citep{2018FrASS...5...25R} is designed to do exactly this science and, if launched, would provide detailed observations of shock breakout, facilitating a better disentangling of the different asymmetries in this signal.  For shock breakout to truly be used to constrain the supernova engine, we need to obtain detailed observations of a large number of events and couple them to increasingly detailed explosion models.

\subsection{Indirect Probe:  Compact Remnant Properties}

The properties of the compact remnant (both mass and spin distributions) provide a probe of the central engine.  Engines driven by rotational energy (either magnetar or disk engines) require high rotation speeds.  Current observations of pulsar spins suggest that such high rotation speeds are not common~\citep{2006ApJ...643..332F,2013MNRAS.432..967I,2022IJMPD..3130005F}.  However, especially for magnetars, determining the birth spin rates can be difficult.  But it is harder to disguise birth spin rates of black holes which are believed to be low, arguing that the stars that form them have little angular momentum~\citep{2020A&A...636A.104B,2021PhRvD.104h3010R}.  If we extrapolate the low angular momenta from black-hole forming stars down to neutron-star forming stars, we obtain the observed low spin rates (few below 5\,ms) of the observed pulsar distribution.  Current observations strongly suggest that angular-momentum powered engines are rare.  

Other aspects of the compact remnants also can constrain properties of the supernova engine.  Rotationally-driven engines do not predict the mass distributions of compact remnants.  Right now, these engines can not be constrained by the remnant mass distribution.  But our understanding of the convective engine allows scientists to make preliminary predictions of the compact remnant mass distribution.  These initial predictions argued for a distribution of compact remnant masses~\citep{2001ApJ...554..548F}.  At the time, observations predict delta-function mass distributions~\citep{1999ApJ...512..288T} and the ultimate discovery that the masses were more broadly distributed is one of the strong validations of the convective engine.  As the remnant mass distribution becomes increasingly refined, it probes the convective supernova engine even further.  For example, enhancements in our understanding of the convective engine have allowed astronomers to apply uncertainties of the growth time of convection to predict remnant mass distributions~\citep{2012ApJ...749...91F,2022ApJ...931...94F}.  

Asymmetries in the central engine can also be probed by the velocity distribution of compact remnants (assuming the velocities are produced by asymmetric ejecta from large-scale convection;~\citealt{1995PhR...256..117H,1996PhRvL..76..352B,2004ApJ...601L.175F,2013A&A...552A.126W,2017ApJ...837...84J}).  The relative velocities of neutron star and black hole systems can also help determine whether these kicks are produced through asymmetric ejecta or asymmetric neutrino emission~\citep{2006ApJS..163..335F}.  But such measurements are indirect probes of the convective engine and depend sensitively on the assumption that kicks are produce by asymmetric ejecta.  If pulsar velocities are produced by alternate kick mechanisms, e.g. a neutrino kick mechanism~\citep{2006ApJS..163..335F}, such measurements may not tie directly to the convective engine.
 
\subsection{Indirect Probe:  Galactic Chemical Evolution}
\label{sec:gce}

The nucleosynthetic yields from core-collapse supernovae can probe both the inner engine and the propagation of the shock, but they are one of the most indirect probes.  The yields of the innermost ejecta depend sensitively on the electron fraction that is, in turn, set by the nature of the engine (convection or disk properties) and the neutrinos (probing dense nuclear matter and neutrino physics) emitted in the engine.  In fact, deleptonization of the material in this innermost ejecta was one of the major limitations of the core-collapse engine~\citep{1980JPhys..41C..25A} until detailed models found that the neutrinos could reset the electron fraction and avoid the production of neutron-rich isotopes~\citep[e.g.][]{2006NewAR..50..496F}.  

The nature of the explosion can also affect the shock propagation yields.  Most spherically-symmetric explosions produce trajectories (density and temperature evolution with time) that match the adiabatic exponential and power-law evolution assumed in many yield studies~\citep[e.g.][]{2010ApJS..191...66M}.  Asymmetric explosions drastically alter these trajectories producing a wide range of results including shocks that reheat the ejecta~\citep{2020ApJ...895...82V,2021ApJ...921..113S}.  These altered trajectories change the nucleosynthetic yields and detailed measurements of the yields can constrain the nature of the explosion.  

Astronomers have used a variety of probes to measure the nucleosynthetic yields.   Galactic chemical evolution models combine the yields ejected from both winds in low-mass stars and supernova explosions of high-mass stars to compare to the observed abundances in stars~\citep{1995ApJS...98..617T,2020ApJ...900..179K}.  Galactic chemical evolution models use a series of assumptions to produce these models:  yields typically based on approximate models, star formation history, initial mass function, binary effects, etc.  These are typically combined in a single zone Galactic model with simplified assumptions for their mixing and models are just now incorporating more detailed physics~\citep[e.g.][]{2021MNRAS.505.5862W}.  

The uncertainties in these calculations, many of which have yet to be characterized, have limited the constraints Galactic chemical evolution models can place on our understanding of supernovae.  To date, this diagnostic has yet to produce reliable constraints on the supernova engine (which stars explode, their explosion energies or the nature of the asymmetries that probe the supernova engine).  

\subsection{Indirect Probe:  Other Constraints on Nucleosynthetic Yields}

Another promising avenue to constrain nucleosynthetic yields is through observations, particularly of isotope ratios, of presolar stardust grains. Grains that condensed in the ejecta of supernova recorded the isotopic signatures of the region they formed in. Some of these grains have been incorporated into meteorite parent at the start of the Solar System 4.56\,Gyr ago and can today be recovered and analyzed \citep[e.g.,][]{2016ARA&A..54...53N}. In order to use these grains as messengers of the supernova engine, a deeper understanding of their formation location and conditions is required.  Although modeling of dust grains is starting to include results of supernova models~\citep{2013ApJ...776..107S,2015A&A...575A..95S,2019BAAS...51c..66S,2022ApJ...931...85B}, advances in both the incorporation of supernova mixing and dust grain production and destruction are needed to tie these observations to the supernova engine. Even with these detailed models, many of the observations will constrain shell burning features of massive stars more than the supernova engine itself. Detailed mixing models will allow deciphering which isotopic signatures are associated with the supernova engine rather than with shell burning nuycleosynthesis and these signatures need to be subsequently analyzed in presolar stardust grains. This requires close collaboration between the modeling and stardust grain measurement communities.

Because of their limited lifetime, radioactive isotopes probe for a limited number of explosions, a cleaner constraint than full Galactic Chemical evolution.  The improved maps of $^{26}Al$, $^{60}Fe$ and, potentially, r-process yields from satellite missions such as the COmpton Spectrometer and Imager (COSI)~\citep{2019BAAS...51g..98T} will constrain our understanding of supernovae.  If COSI can demonstrate that some radioactive r-process elements are formed in supernovae, this will constrain both the supernova engine and the physics driving the explosions.  $^{26}$Al and $^{60}$Fe are less constraining of the engine itself.  Even though large amounts of $^{26}$Al and $^{60}$Fe are produced in the supernova explosion~\citep{2019MNRAS.485.4287J}, the yields of these isotopes depend sensitively on the conditions in the shell burning and much more work is needed to disentangle the conditions in the star and the explosion mechanism.

In this paper, we study two measurements of these yields:  supernova remnants (Section~\ref{sec:SNR}) and gamma rays (Section~\ref{sec:gray}).  Galactic chemical evolution (see Sec.~\ref{sec:gce}) is a more indirect diagnostic.

\section{Mixing and Model Grid}
\label{sec:grid}

Asymmetries in the explosion will cause the elements in the supernova ejecta to mix together and can cause dramatic deviations from the ``onion-skin" picture of stellar and supernova element distributions.  The new studies in this paper are primarily based on a grid of artificially-mixed supernova models.  For our calculations, we leverage the grid of 1-dimensional models, described in \cite{2018ApJ...856...63F} and \cite{2020ApJ...890...35A}, spanning a range of supernova explosion energies produced by altering the convective engine for three stellar progenitors (15, 20, and 25\,$M_\odot$ stars).  For each of these progenitor stars, we pick 2 explosions, a low and high-energy explosion.  Each of these progenitors/explosions produces a different amount of $^{56}$Ni.  In the more massive progenitors, fallback is likely to occur~\citep{2006NewAR..50..492F,2009ApJ...699..409F} and, for our 20\,$M_\odot$ and 25\,$M_\odot$ progenitors, even the strong explosion does not eject much $^{56}$Ni.  Nonetheless, such models can produce luminous supernovae, especially if shock heating is strong.  Table~\ref{tab:explosions} shows the energies and $^{56}$Ni yields for each of our models.

\begin{table}[ht]
    \begin{center}
    \begin{tabular}{lcccc}
    \hline
    Model & M$_{\rm prog}$ & E$_{\rm exp}$ & M$_{\rm ejecta}$ & M$_{Ni}$ \\
    & $M_\odot$ & $10^{51}$\,erg & $M_\odot$ & $M_\odot$  \\
    \hline
    M15E0.9 & 15 & 0.9 & 13.3 & 0.056 \\
    M15E2.5 & 15 & 2.5 & 13.5 & 0.105 \\
    M20E1.5 & 20 & 1.5 & 9.0  & 0.0 \\
    M20E4.3 & 20 & 4.3 & 11.2 & 0.0 \\
    M25E1.6 & 25 & 1.6 & 10.7 & 0.0 \\
    M25E7.5 & 25 & 7.5 & 13.0 & 0.0 \\
    \hline
    \end{tabular}
    \caption{Key properties of our supernova explosions:  progenitor mass, explosion energy, ejecta mass and $^{56}$Ni mass.  These explosions, from \cite{2018ApJ...856...63F} are driven by artificial energy deposition within a convective engine where both the power and duration of the engine are modified to produce a variety of explosion energies.  The model name is based on the progenitor mass and explosion energy.}
    \end{center}
    \label{tab:explosions}
\end{table}

The focus of this paper is to study a few supernova observables (gamma-ray emission, supernova spectra and light-curves across the UV, optical and IR bands, and supernova remnants).  Asymmetric explosions cause extensive mixing and we can use this mixing to probe the explosion.  The outward mixing of the $^{56}$Ni is one of the clearest effects on these observables.  For two decades, scientists have studied the evolution of asymmetric explosions on the distribution of the $^{56}$Ni within the ejecta~\citep{2003ApJ...594..390H,2005ApJ...635..487H,2012ApJ...755..160E,2013ApJ...773..161O, 2017ApJ...842...13W, 2019A&A...624A.116U,2020ApJ...895...82V,2020ApJ...888..111O,2021ApJ...914....4U}.  However, most of these studies were limited to a few 3-dimensional calculations based on the current status of supernova explosions.  For example, \cite{2003ApJ...594..390H} used the explosion models from \cite{2004ApJ...601..391F} to guide a set of bimodal and unimodal models to explain qualitative features observed in SN 1987A.  \cite{2017ApJ...842...13W} followed an explosion from a first-principles engine calculation out through late times.  However, no engine calculation has, as yet, captured all of the physics (neutrino processes, equation of state, magnetic fields, etc.) or resolved the convection and convective seeds in the explosion~\citep{2022ApJ...931...94F}.   Added to the uncertainties in the pre-collapse progenitors~\citep[see, for example][]{2021ApJ...921...28F}, although it is possible to study trends in the behavior of supernova explosions, we are far from achieving exact solutions or quantitative estimates of the extent of the mixing.

Guided by these simulations, we parameterize the mixing driven by the convective engine.  For each of our progenitor/explosion energies, we describe the mass coordinate and the extent of our mixing.  For a given extent of the mixing, the mixing mass coordinate ($M_{\rm mix}$), we calculate the average abundance fractions ($\left< f^i \right>$:
\begin{equation}
\left< f^i \right> = \int_0^{M_{\rm Mix}} f^i(m) dM / M_{\rm mix}
\end{equation}
where $f^i(m)$ is the abundance fraction as a function of mass coordinate.  The extent of the mixing is determined by the mixing fraction ($f_{\rm mix}$) where our final abundances within the $M_{\rm mix}$ are given by:
\begin{equation}
    f^i_{\rm final}(m) = f_{\rm mix} \left< f^i \right> + (1-f_{\rm mix}) f^i
\end{equation}
This prescription conserves the total abundances in the explosion but allows us vary the extent of the mixing.  For example, if $M_{\rm mix}$ were set to the total ejecta mass and $f_{\rm mix}=1$, the ejecta would be thoroughly mixed with a uniform composition.  Most explosion calculations suggest that the innermost ejecta is mixed out through part of the star with only a fraction of the innermost ejecta getting mixed outward:  e.g. $M_{\rm mix}$ set to the helium core mass and $f_{\rm mix}=0.25$.  In such a moderate mixing scenario, most of the $^{56}$Ni remains in the innermost ejecta, but a small fraction is ejected outward.  Because there is little/no $^{56}$Ni produced in our more massive progenitors, we focus most of our studies of mixing on the 15\,$M_\odot$ progenitor.  For this study, our suite of explosion models and their mixing parameters are listed in Table~\ref{tab:mixmodels}.

\begin{table}[ht]
    \begin{center}
    \begin{tabular}{lcc}
    \hline
    Model & $M_{\rm mix}$ & $f_{\rm mix}$ \\
    & $M_\odot$ & \\
    \hline
    M15E0.9M0.25f0.25 & 0.25 & 0.25 \\
    M15E0.9M0.25f1 & 0.25 & 1.0 \\
    M15E0.9M4f0.25 & 4 & 0.25 \\
    M15E0.9M4f1 & 4 & 1.0 \\
    M15E0.9M13.3f0.25 & 13.5 & 0.25 \\
    M15E0.9M13.3f1 & 13.5 & 1.0 \\
    M15E2.5M0.25f0.25 & 0.25 & 0.25 \\
    M15E2.5M0.25f1 & 0.25 & 1.0 \\
    M15E2.5M4f0.25 & 4 & 0.25 \\
    M15E2.5M4f1 & 4 & 1.0 \\
    M15E2.5M13.5f0.25 & 13.5 & 0.25 \\
    M15E2.5M13.5f1 & 13.5 & 1.0 \\
    M20E1.5M0.25f0.25 & 0.25 & 0.25 \\
    M20E1.5M9.0f0.25 & 9.0 & 0.25 \\    
    M20E4.3M0.25f0.25 & 0.25 & 0.25 \\
    M20E4.3M11f0.25 & 11.2 & 0.25 \\    
    M25E1.6M0.25f0.25 & 0.25 & 0.25 \\
    M25E1.6M11f0.25 & 10.7 & 0.25 \\    
    M25E7.5M0.25f0.25 & 0.25 & 0.25 \\
    M25E7.5M13f0.25 & 13.0 & 0.25 \\      
    \hline
    \end{tabular}
    \caption{Mixed models and their mixing parameters with the extent of the mixing and the mixing fraction.  The filename is based first on the progenitor mass (M15, M20, M25), the supernova explosion energy (e.g. E0.9, E2.5), the extent of the mixing (e.g. M0.25, M13.5) and, finally, the mixing fraction (f0.25,f1.0).}
    \end{center}
    \label{tab:mixmodels}
\end{table}

Our mixing model implementation is not so different than models used in the past to study gamma rays~\citep{1988ApJ...329..820P} and nucleosynthetic yields~\citep{2002ApJ...565..385U,2007ApJ...660..516T,2009ApJ...690..526T}. The different approaches are partially set by the explosion models used and by the observational focus.  The \cite{2002ApJ...565..385U,2007ApJ...660..516T,2009ApJ...690..526T} include a method to include fallback.  Fallback was studied in the suite of models used in this paper~\cite{2018ApJ...856...63F,2020ApJ...890...35A} and we used models to produce a range of results, but alternative methods of fallback~\citep[e.g.][]{2007ApJ...660..516T} would produce slightly different quantitative results, but the trends (and ultimate conclusions) from different mixing models will be the same independent on the exact details of the mixing model. 

It is important to understand that this mixing does not capture all of the features of the asymmetric explosions.  Capturing all the features of asymmetric shocks from the convective engine, or jet models, and their effect on nuclear burning and their combined effect on supernova observables requires 3-dimensional models.  However, this parameterized study will give us a first pass at the importance of the effects of the asymmetries produced in the different explosive engines.

\section{Supernova Remnants}
\label{sec:SNR}

Supernova ejecta produce remnants that are a powerful probes of the engine.  The expansion of these remnants allow astronomers to spatially resolve the remnant, providing a unique window not only into the ejecta composition but also the distribution (and asymmetries) of this ejecta.  Combining a broad set of observations from radio to gamma ray, scientists have been able to probe supernova explosion properties including the explosion energies and asymmetries.  Supernova remnants have provided some of the most direct support for the convective engine.  The difficulties that current 1-dimensional models have in explaining the yields in supernova remnants demonstrates the power of remnant observations to constrain properties of the supernova and its progenitor (Braun et al. 2023, in preparation).  But we are currently limited by what we can detect (sensitivity) and the spatial resolution of the instruments.

But using observations of supernova remnants suffers from a few complicating factors.    As the blastwave propagates through the material surrounding the star, it decelerates, sending a reverse shock through the supernova ejecta.   Most observations of the supernova ejecta are limited to material heated by this reverse shock and, in many supernova remnants, we are only observing part of the ejecta.  In addition, the circumstellar medium, much of which is driven by mass outflow from the star (from winds or binary interactions) can often deviate from spherical symmetry.  These asphericities in the circumstellar medium can drastically shape the features and instabilities of the outflow~\citep{1996ApJ...472..257B}.  This emission depends on the characteristics of both the supernova explosion and the circumstellar medium~\citep{2013arXiv1305.4137E}.  We can not take for granted that the features in the supernova remnant are caused by asymmetries in the supernova explosion.    By mapping the distribution of $^{44}$Ti, which is produced in the central engine, scientists were able to demonstrate a multi-lobe asymmetry in the explosions, as predicted by the convective engine~\citep{2014Natur.506..339G,2017ApJ...834...19G}.  After observing the $^{44}$Ti features of Cassiopeia A~\citep{2014Natur.506..339G}, it became clear that the jet-like structures~\citep{2004ApJ...615L.117H} in that remnant were not caused by asymmetries in the supernova explosion (e.g. jet) but perhaps were produced by explosion/circumstellar medium interactions.

The initial conditions are not the only difficulty in using remnants to measure the abundances and abundance distributions from supernovae (and hence constrain the engine properties).  As the supernova blast wave moves outward, many equilibrium assumptions begin to break down.  For example, remnant shocks are typically in the ``collisionless regime" where the ion and electron distributions can deviate from a thermal Maxwellian~\citep[e.g.][]{1979ApJ...234L.195P, 2022ApJ...929....7V,1981AdSpR...1m..71C,1983PhDT.........2N}.  In addition, the atomic level states, typically set by collisional and radiative processes can deviate from equilibrium and even steady-state solutions~\citep{2010ApJ...725.1476P,2014PhDT.......477B,2015A&A...579A..13V,2017ApJ...851...12R,2020ApJ...903....2R,2021MNRAS.504..583S}.  Inferring abundances in such conditions can be challenging.  All of these deviations from equilibrium, including cosmic ray production, are actively studied by the remnant community~\citep{2020ApJ...903....2R,2021NJPh...23e3010Y,2021MNRAS.504..583S,2022ApJ...929....7V}.  

Another complicating factor lies in the fact that although spectral identifications are ideally suited to determine elemental distributions, they are less powerful at differentiating isotopes.  Many elements synthesized in the central engine exist as stable elements in the star.  At solar metallicity, these stable elements can dominate the signal.  For example, in the \cite{1995ApJS..101..181W} yield database, 15-25\% of the iron in solar-metallicity explosions is not produced in the star or the explosion, but instead was there pre-explosion from the solar distribution of the gas forming the star.  In the \cite{2020ApJ...890...35A} models that consist of a much broader range of explosion energies and a more realistic implementation of the explosion to include fallback, this fraction ranges from 8-100\%, where the 100\% models corresponds to a supernova where all of the synthesized iron falls back onto the neutron star.  To unambiguously probe the supernova engine, we must focus on elements which are only present or predominantly present in the innermost ejecta.  

\begin{figure}[ht]
\includegraphics[scale=0.45,angle=0]{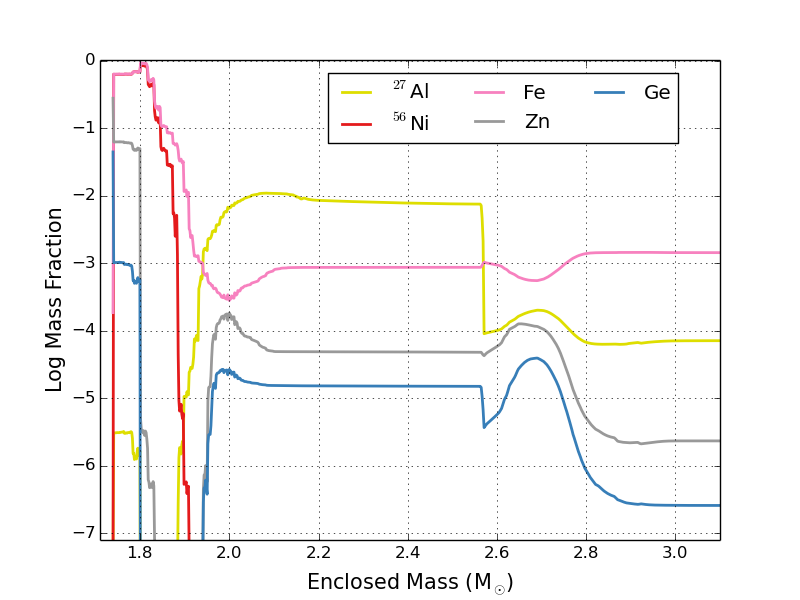}
\includegraphics[scale=0.45,angle=0]{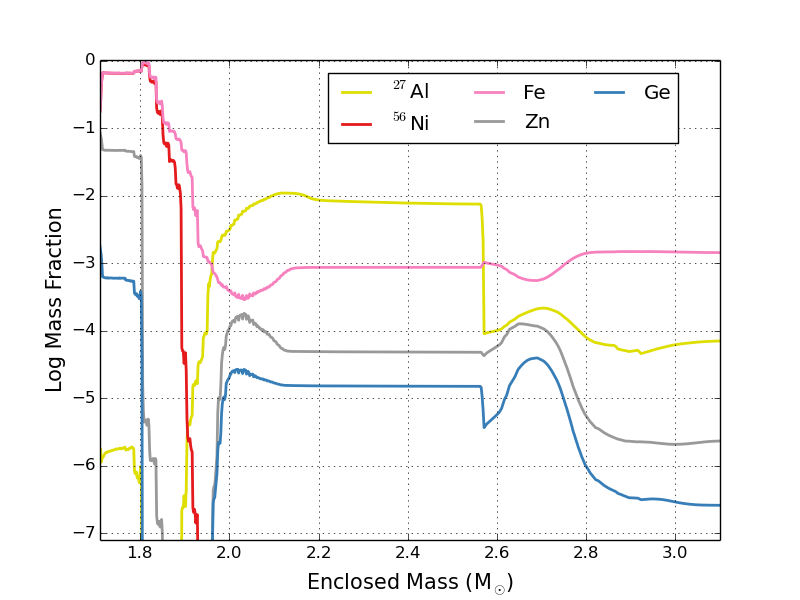}
\caption{Distribution of select elements in two supernova explosions of a 15\,M$_\odot$ solar-metallicity progenitor with energies of $0.9\times 10^{51} {\rm erg}$ (left) and $2.6\times 10^{51} {\rm erg}$ (right).  Although $^{56}$Ni probes only the innermost ejecta, a large amount of its ultimate decay product, $^{56}$Fe, exists in the solar-metallicity material that forms its stars.  Ge and Zn are also produced in the central engine and less contaminated by solar abundances.} 
\label{fig:mdist}
\end{figure}

Detailed modeling will help us to identify the best probes of this central engine.  For example, Figure~\ref{fig:mdist} shows the distribution of elements as a function of enclosed mass from the explosion of a 15M\,$_\odot$ star with two different explosions energies.  The weaker explosion has more fallback, and the iron produced in the explosion is reduced by this fallback (this is seen in the drop to zero of all yields at the innermost region in enclosed mass).  In this case, the iron in the solar metallicity gas forming the star dominates the iron in the remnant.  For sufficiently-strong explosions, the iron distribution remains an excellent probe of the explosion asymmetry.  But other elements are less dominated by the metals in the gas.  Although the zinc and germanium elements exist in the solar abundance pattern (and hence in the star upon formation), $90-99$\% of the zinc and germanium mass ejected in a supernova is formed during stellar evolution or the supernova explosion itself\footnote{The difference between iron yields and germanium and zinc yields is that thermonuclear supernovae produce most of the iron in the Galaxy whereas core-collapse supernovae are more dominant producers of germanium and zinc}.  The amount of these elements produced is sensitive to the neutrino physics because it sets the electron fraction and hence they can be used to probe this physics.  In \cite{2020ApJ...890...35A}, zinc to iron ratio varies from $0.005-0.4$ and the germanium to iron ratio varies from $0.01-0.3$.  In addition, their spectral lines are observed in the UV and X-ray, leveraging the more sensitive telescopes in these energy bands.

\begin{figure}[ht]
\includegraphics[scale=0.45,angle=0]{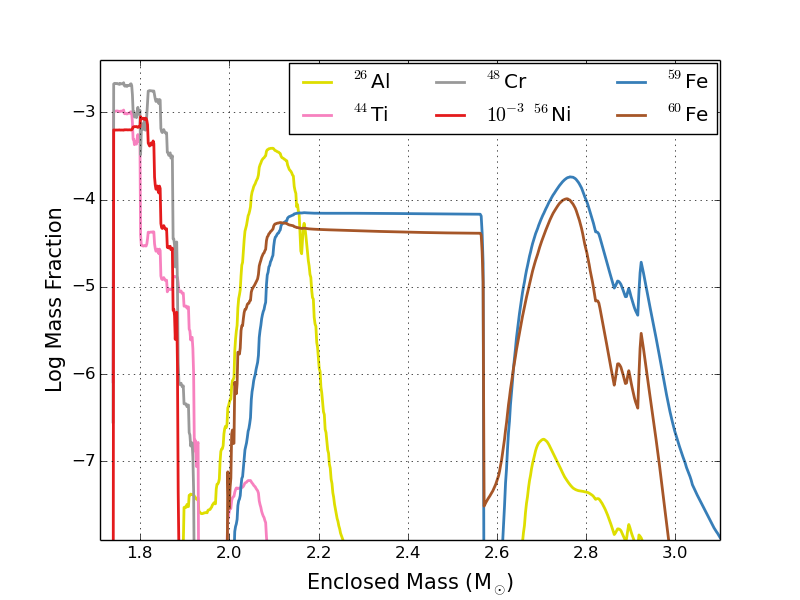}
\includegraphics[scale=0.45,angle=0]{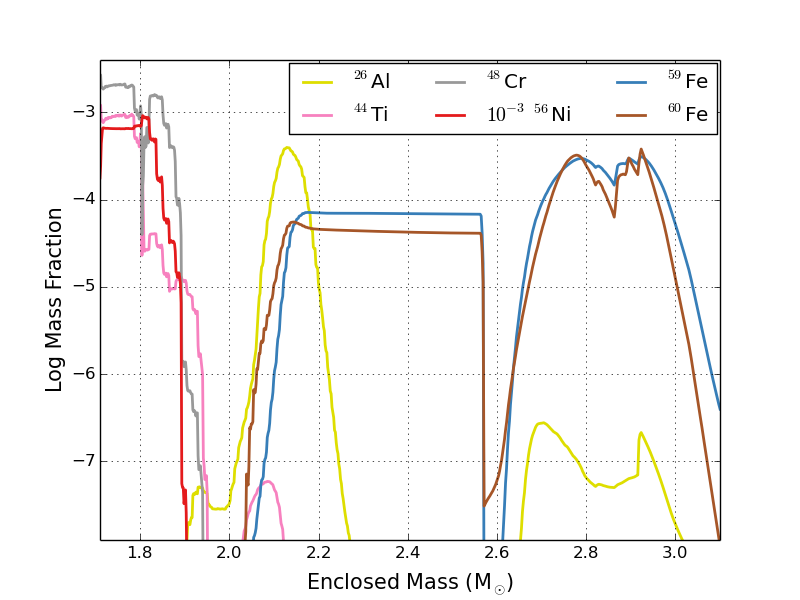}
\caption{Distribution of select radioactive isotopes in two supernova explosions of a 15\,M$_\odot$ solar-metallicity progenitor with energies of $0.9\times 10^{51} {\rm erg}$ (left) and $2.6\times 10^{51} {\rm erg}$ (right).  $^{44}$Ti, $^{48}$Cr, and $^{56}$Ni all probe the supernova engine.  $^{26}$Al probes the boundary layer between silicon and carbon/oxygen shells.  And neutron-rich iron isotopes, e.g. $^{59,60}$Fe, are produced in both the carbon/oxygen and helium shells.  Although $^{26}$Al, $^{59,60}$Fe are all produced in the shell burning prior to collapse, the explosion drives further production and destruction.} 
\label{fig:mdistrad}
\end{figure}

One way to avoid many of these issues is to focus on emission from the decay of radioactive isotopes.  A growing number of studies have shown the correlation between the nature of the explosive engine and these yields~\citep{2017ApJ...842...13W,2019ApJ...886...47S,2020ApJ...890...35A}.  This emission does not require shock heating and is much less sensitive to the circumstellar medium and out-of-equilibrium effects.  Because these isotopes decay, they are not in the solar abundance pattern and the only isotopes observed are synthesized either in the star or the supernova explosion.  Radioactive isotopes produced in the central engine are ideal probes of the nature of this engine.  From Figure~\ref{fig:mdistrad}, we see that $^{44}$Ti is a good tracer of the innermost ejecta and {\it NuSTAR} observations of the Cassiopeia A supernova remnant~\citep{2014Natur.506..339G,2017ApJ...834...19G} provide the strongest evidence case demonstrating that the convective engine produces at least some supernova explosions. Unfortunately, the only radioactive isotope produced in the central engine in quantities that can be observed and with a half-life that allows it to be observed in remnants is $^{44}$Ti, and its 60\,y half-life limits observability to a narrow window.  Combined with the difficulty in line detection, it will not be easy to map out additional remnants like Cassiopeia A through radioactive decay.  

\begin{figure}[ht]
\includegraphics[scale=0.45,angle=0]{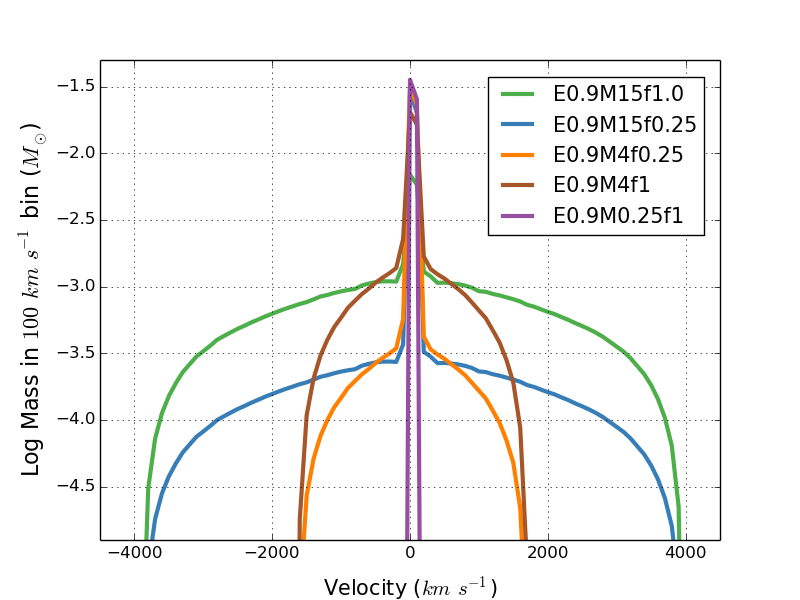}
\includegraphics[scale=0.45,angle=0]{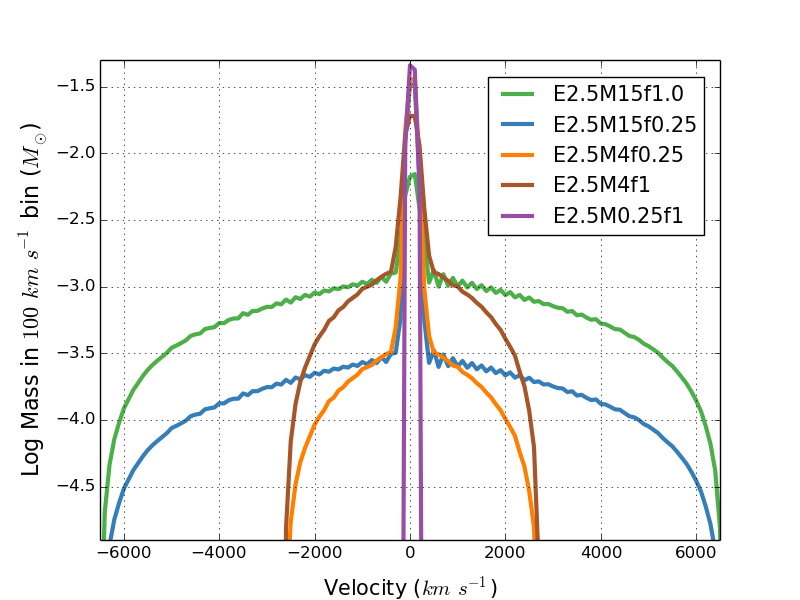}
\caption{Mass of $^{44}$Ti as a function velocity for a variety of our artificially mixed models.  Because of the mass distribution as a function of velocity, even with a fully mixed solution, there is a peak in the mass at low velocities.  But the mass can be spread out to high velocities.} 
\label{fig:velexp}
\end{figure}

Instead, most probes of mixing using radioactive isotopes will focus on measuring the lines themselves.  Figure~\ref{fig:velexp} shows the velocity profiles for a few of the explosion models in our study.  This figure compares the velocity profiles of both $^{56}$Ni and iron from a subset of our mixing models.  The observed velocity profile in a supernova remnant can help provide insight into the extent of the mixing in the supernova itself.  If we can observe the full extent of the line profiles (depending upon the amount of mixing, these line wings can be orders of magnitude dimmer than the peak line), we can use the line widths to probe the mixing of the elements.  If we focus on those elements predominantly produced in the engine, we can measure the extent of the engine asymmetry.

\begin{figure}[ht]
\includegraphics[scale=0.43,angle=0]{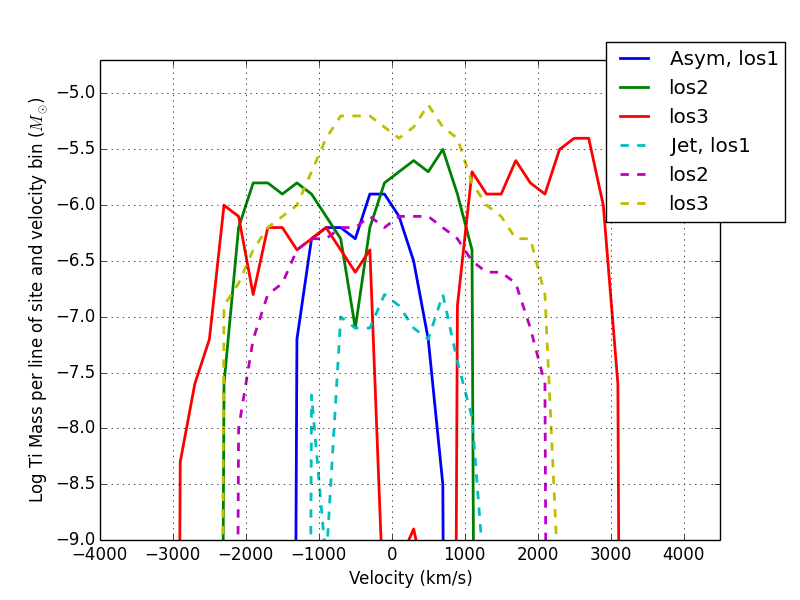}
\includegraphics[scale=0.43,angle=0]{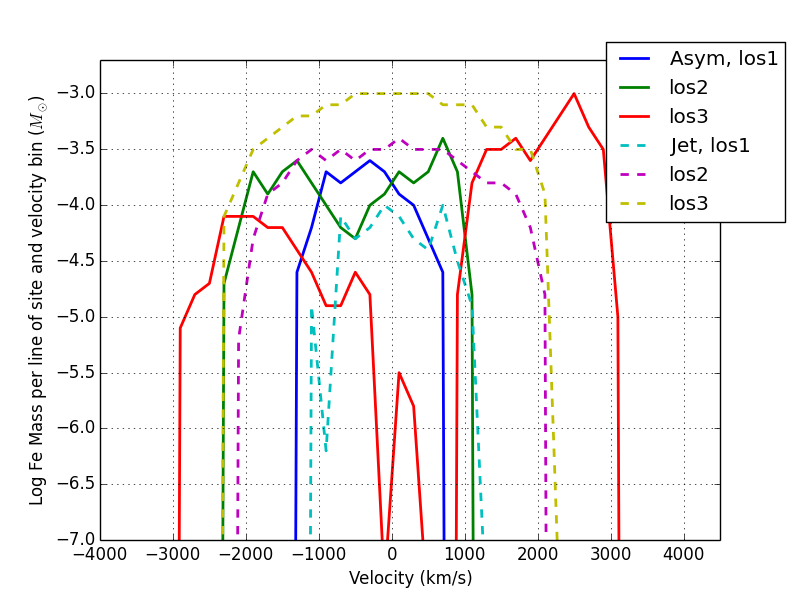}
\caption{Distribution of $^{44}$Ti (left) and $^{56}$Ni (right) in two different 3-dimensional explosions each along three different lines of site.  Because of the more asymmetric explosion velocities, there is less of a peak in mass at really low velocities.  Clumps of high moving ejecta from the supernova engine can escape with higher velocities.} 
\label{fig:vel3D}
\end{figure}

Figure~\ref{fig:velexp} uses the artificially mixed models from our current study.  In these models, we assume that the abundance {\it fraction} is mixed.  The mass or density distribution versus mass produces a distribution that, even if the material is uniformly mixed, places a large amount of mass at low velocities.  We have also studied a variety of models in 3-dimensions.  The nature of a 3-dimensional explosion can cause material to plow through the ejecta, producing clumps of material at much higher velocities.  For these models, the peak at low velocities can be much less pronounced than our simple mixing models.  Figure~\ref{fig:vel3D} shows the $^{44}$Ti velocity distribution of a set of 3-dimensional explosions with varying levels of asymmetries from the convective or jet-driven engines~\citep{2012ApJ...755..160E,2020ApJ...895...82V}.  Additional studies of such asymmetries are necessary to determine how well these observations can differentiate between different levels of asymmetries and their engines.

By observing these distributions, either by mapping out supernova remnants or measuring line broadened profiles, astronomers can probe the level of asymmetry in the explosion.  We have already discussed how {\it NuSTAR} observations of the $^{44}$Ti distribution of the Cassiopeia A remnant provides some of the strongest evidence for the convective engine~\citep{2014Natur.506..339G,2017ApJ...834...19G,2021Natur.592..537S}.  With its broad energy range from 0.1-150\,keV{\it HEX-P}~\citep{2018SPIE10699E..6MM} has the potential take the {\it NuSTAR} observations to the next level, providing further constraints.  {\it JWST} observations of unshocked iron in the Cassiopeia A remnant~\citep{2021jwst.prop.1947M} will allow astronomers to conduct detailed comparisons between $^{56}$Ni (which decays to iron) and $^{44}$Ti production in the central engine.  High-resolution X-ray spectroscopy with the soon-to-be-launched XRISM satellite~\citep{2020arXiv200304962X}  will map the spectral lines with an unprecedented resolution needed to constrain progenitors.  High-energy missions such as COSI~\citep{2019BAAS...51g..98T}, Strobe-X~\citep{2019arXiv190303035R}, and ASCENT~\citep{2019arXiv190303035R} will measure line profiles that, coupled with models such as those in Fig.~\ref{fig:vel3D}, will allow us to probe the supernova engine for a wider set of supernova remnants.  As we extend the elements probed (e.g. Ge, Zn), we can also study different aspects of the supernova progenitor and its engine.

\section{Supernova Lightcurves and Spectra}
\label{sec:lc}

For SN 1987A, the $^{56}$Ni produced in the innermost ejecta was mixed out in the star, causing it to achieve faster velocities.  These faster velocities broaden the line profiles and late-time observations of broad infra-red lines comprised part of the evidence file of mixing in SN 1987A~\citep{1990MNRAS.242..669S,1990ApJ...360..257H}.  These observations demonstrated the strength of late-time spectral features in proving supernova mixing.  But such late-time observations are difficult to obtain and were only made possible by the proximity of SN 1987A.  For most SNe, such observations do not exist.  Depending on the mixing, there may be other observable features in the UV, optical, and IR that can probe the mixing.  In this section, we compare light-curve and spectra calculations of our grid of models to determine observable features in this emission that can be used to help probe mixing in supernovae.

For our spectra and light-curve calculations, we use the {\it SuperNu} Monte Carlo method that couples both Implicit Monte Carlo Methods~\citep{1971JCoPh...8..313F} with Discrete Diffusion Monte Carlo~\citep{2012JCoPh.231.6924D} in optically thick regions~\citep{2013ApJS..209...36W,2014ApJS..214...28W}.  Heating from radioactive decay assume in-situ energy deposition from electrons/ions and a gray-opacity gamma-ray transport implementation for gamma-ray deposition.  This gamma-ray transport has been tested against multi-group opacity gamma-ray methods~\citep{2003ApJ...594..390H}, achieving good agreement in the amount and disposition of the energy deposition~\citep{2017ApJ...845..168W}.  Although the calculations are transport-only and assume a homologous outflow, we have included the effects of shock heating by adding a small fraction of kinetic energy (up to 10\%) to the thermal energy in the explosion. Our grid of models include a set of models with this enhanced shock heating, discussed below.

\begin{figure}[ht]
\includegraphics[scale=0.45,angle=0]{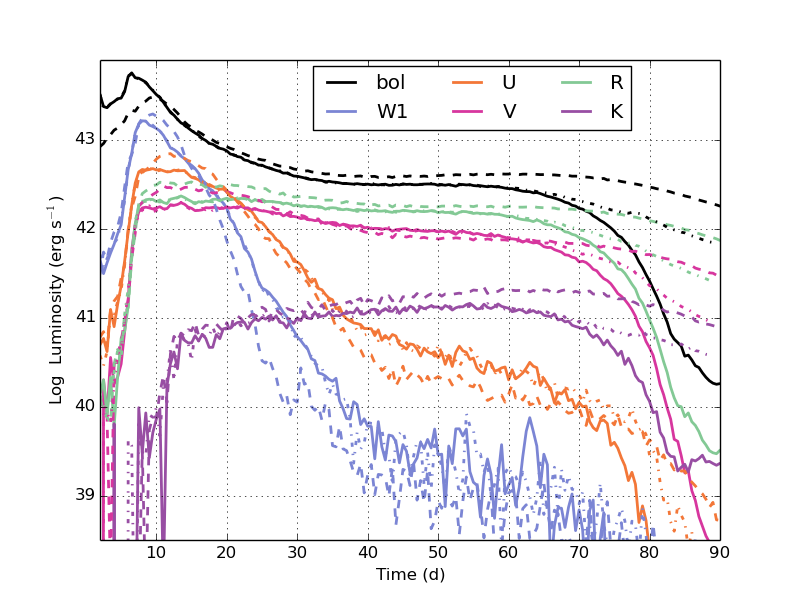}
\includegraphics[scale=0.45,angle=0]{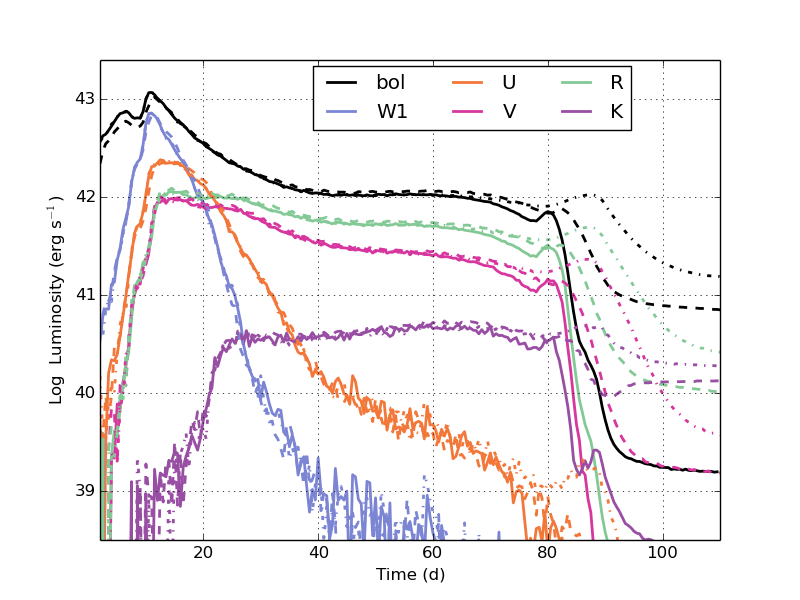}
\caption{U-, V-, R-, K- and Swift W1-band light-curves for both our weak (left) and strong (right) supernova explosions from our 15\,M$_\odot$ progenitor.  The solid line shows minimal mixing limited to the inner 0.25\,M$\odot$ (model M15m0.25f1.0).  The dashed line shows partial mixing throughout the extent of the star (M$_{\rm mix}\approx13$, f$_{\rm mix}=0.25$) and the dot-dashed line shows the full mixing out through the helium core (M$_{\rm mix}\approx4$, f$_{\rm mix}=1.0$).} 
\label{fig:lc15}
\end{figure}

Figure~\ref{fig:lc15} shows the U-, V-, R-, K- and Swift W1-band light-curves for both our weak and strong explosions of our 15\,M$_\odot$ progenitor for minimal, moderate and strong mixing.  These models assume that the shock heating is set to the value produced in our hydrodynamic explosion models.  Because shock heating dominates the energy source, these type II-plateau supernovae are not affected much by the mixing until after the peak and plateau phases.  At late times, as the photosphere moves through the star and the initial shock heating becomes less important, the additional heating from radioactive decay does play a role.  At these late times, the effects of mixing from the central engine alter the light-curves.

\begin{figure}[ht]
\includegraphics[scale=0.45,angle=0]{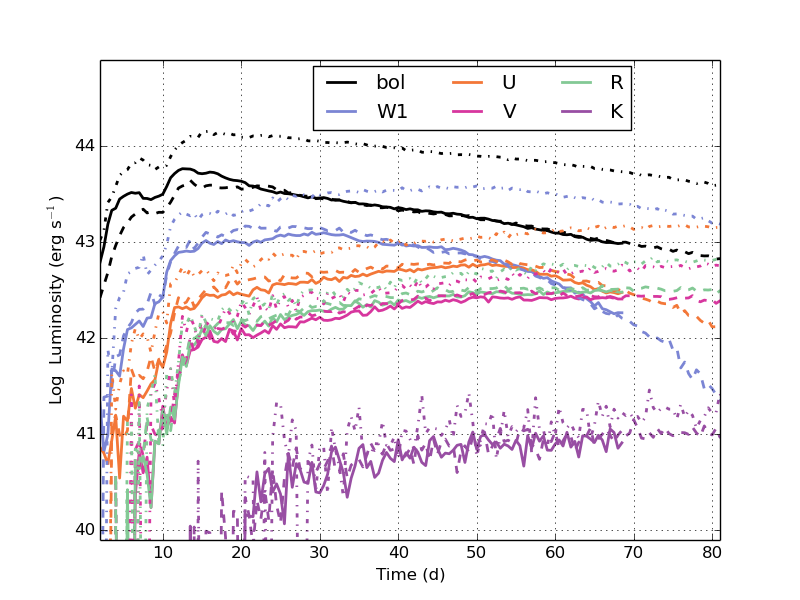}
\includegraphics[scale=0.45,angle=0]{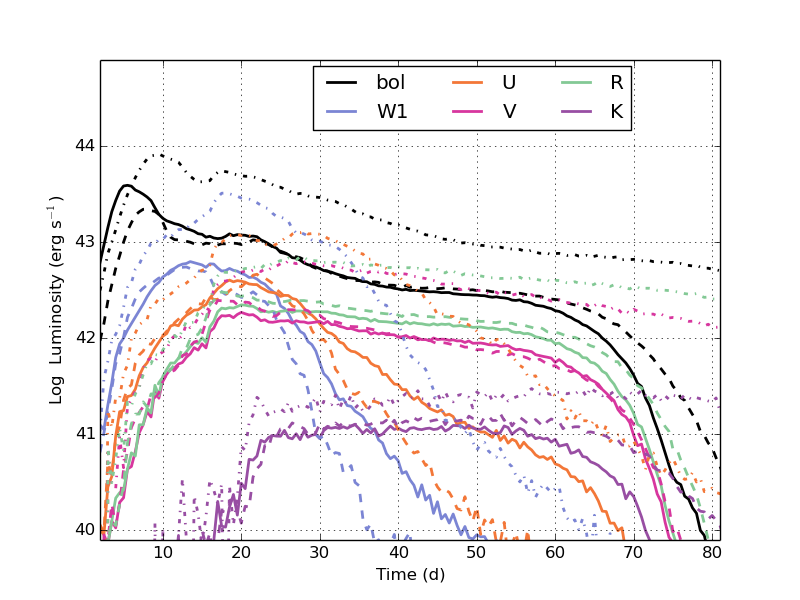}
\caption{U-, V-, R-, K- and Swift W1-band light-curves for both our weak (left) and strong (right) supernova explosions from our 25\,M$_\odot$ progenitor.  The solid line shows minimal mixing limited to the inner 0.25\,M$_\odot$ (model M25m0.25f1.0).  The dashed line shows partial mixing throughout the extent of the star (M$_{\rm mix}\approx13$, f$_{\rm mix}=0.25$).  The dot-dashed line shows this same mixed model but with strong shock heating where 10\% of the kinetic energy is converted to thermal energy.} 
\label{fig:lc25}
\end{figure}

Figure~\ref{fig:lc25} shows the U-, V-, R-, K- and Swift W1-band light-curves for both our weak and strong 25\,M$_\odot$ for minimal and strong mixing.  Because very little $^{56}$Ni is ejected in both our weak and strong 25\,M$_\odot$ progenitor, the only affect of mixing is to alter the opacities in the outflow and the net affect on the light-curve is minimal.  But increasing the amount of shock heating can dramatically alter the light-curve.  Even a small amount of additional shock heating as the shock progresses through the inhomogeneous wind medium can drastically alter the light-curve.  In the UVOIR bands, the emission is much more sensitive to shock heating than it is to the mixing of $^{56}$Ni into the outer layers of the ejecta.

\begin{figure}[ht]
\includegraphics[scale=0.45,angle=0]{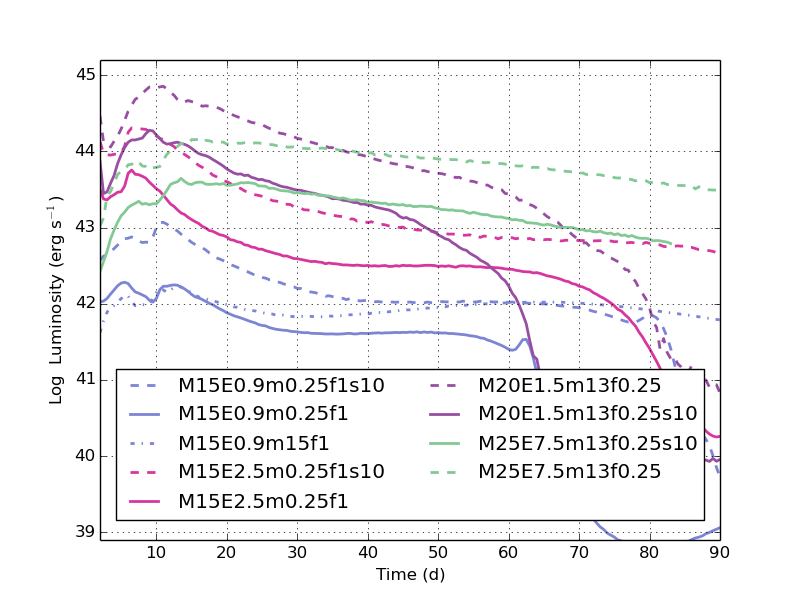}
\caption{Bolometric light-curves for a sampling of our light-curve models varying progenitor mass and shock heating (with a few models varying explosion energy and mixing).  The breadth of the light-curves shows the sensitivity of this emission to shock heating.} 
\label{fig:lcbol}
\end{figure}

To compare the relative effects of progenitor, mixing and shock heating on the light-curves, Figure~\ref{fig:lcbol} shows the bolometric light-curves for a wide set of models including models with no and extreme shock heating conditions.  Shock heating can have a dramatic effect on the light-curve and the extent of this mixing can easily mask a mixing observation.  Although this limits how well UVOIR light-curves can probe mixing, it does mean that this emission is ideally suited to probing shock heating.  Combined with other observations, shock breakout for example~\citep{2020ApJ...898..123F}, UVOIR emission is ideally suited to probing the nature of the circumstellar medium and stellar mass loss.

\begin{figure}[ht]
\includegraphics[scale=0.45,angle=0]{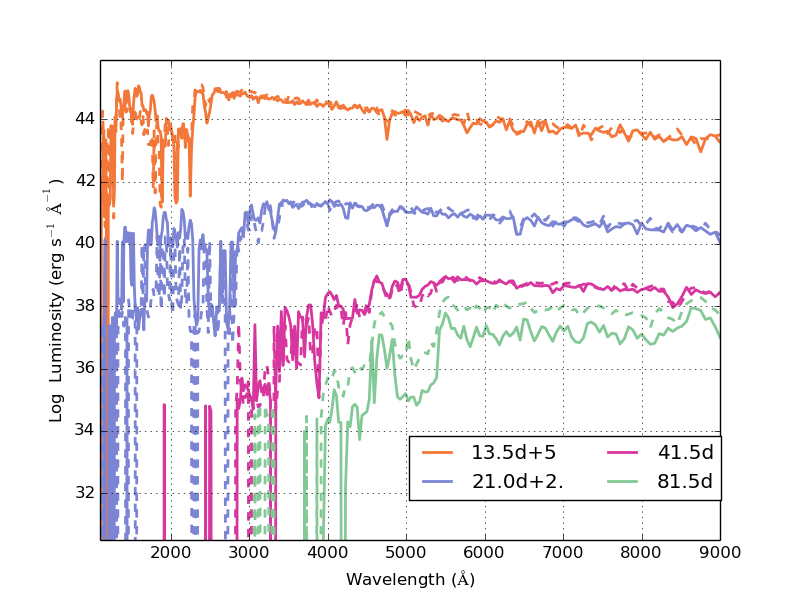}
\includegraphics[scale=0.45,angle=0]{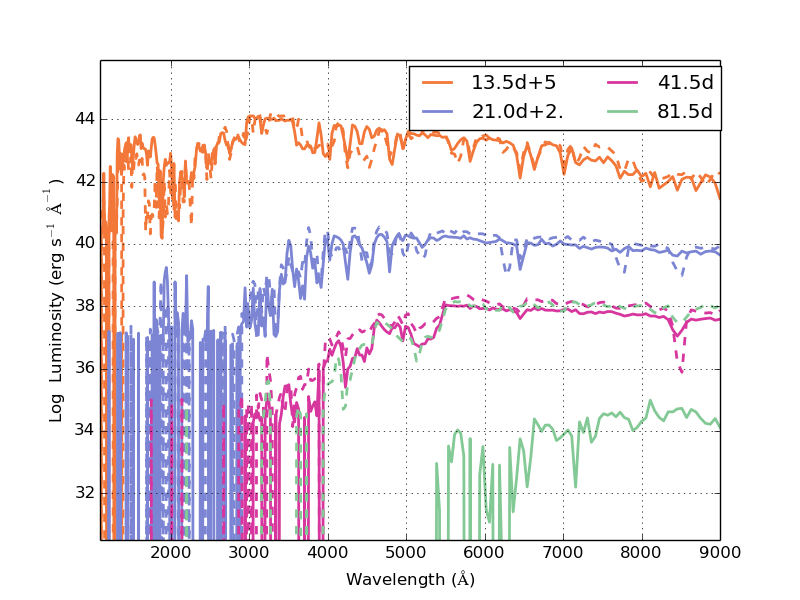}
\caption{Spectral emission (${\rm erg \, s^{-1} \AA^{-1}}$) as a function of wavelength ($\AA$) for our strong (left) and weak (right) explosions of a 15\,M$_\odot$ progenitor.  These models include both low-mix solution with $M_{{\rm mix}}=0.25, f_{{\rm mix}}=1.0$ (solid) and fully mixed with $M_{\rm mix}=13, f_{\rm mix}=1.0$ solutions.  The primary feature of the fully mixed solutions is the appearance of heavy metal lines in the early-time UV spectra.   At late times, the $^{56}$Ni-heating in the low energy explosion alters the spectra dramatically as it did for the band emission.  However, this radioactive heating can be mimicked by shock heating.} 
\label{fig:spectra}
\end{figure}

Spectra can also place constraints on the outward mixing of the elements.  As we discussed at the beginning of this section, the width of the iron lines in the late-time spectra of SN 1987A was part of the evidence file for mixing in supernovae.  Unfortunately, the emission can drop dramatically at late-times and these observations become increasingly difficult.  In Figure~\ref{fig:spectra}, we compare the spectra from highly-mixed models (M15E0.9M13.3f1, M15E2.5M13.5f1)
and low-mixing models (M15E0.9M0.25f1, M15E2.5M0.25f1) at early times (from before peak and throughout the plateau phase).  With high resolution, line broadening will be able to distinguish some features.  The presence of line features from Si-group and Fe-group elements in the early time spectra (primarily in the UV) is a clear indication of extensive mixing.  The large difference between the low-mix and high-mix solutions with our low energy explosion arises because, at these times, $^{56}$Ni heating starts to alter the light-curves in these models (as seen in our band light-curves).  Also note that these comparisons show extreme results (homoegeneously mixed models for our highly-mixed solutions) and the amount of heavy elements in the spectra will be less for any realistic model.

\section{Transient X rays and Gamma Rays}
\label{sec:gray}

The early rise in the MeV gamma-ray emission from SN 1987A~\citep{1989PAZh...15..291S,1990AdSpR..10b..47S,1990AdSpR..10b..55F} was one of the key observations driving the development of the convective engine~\citep{1988ApJ...329..820P} and implementing the explosions produced by the convective engine not only explained the early rise time of the gamma-ray emission but could explain the observed red-shifted line features~\citep{2003ApJ...594..390H,2005ApJ...635..487H}.  Combined with the {\it NuSTAR} images of Cassiopeia A (we discussed these X-ray/gamma-ray detections in the section~\ref{sec:SNR}), these observations form the foundation of the argument for asymmetries in the supernova engine.

For our study of the gamma-ray emission, we use the 3-dimensional {\it Maverick} code used in past studies and verified against a a broad suite of 1-dimensional gamma-ray transport methods~\citep{2003ApJ...594..390H,2004ApJ...613.1101M,2005ApJ...635..487H}.  With this code, we model the same suite of explosions studied with our UVOIR light-curves.  

\begin{figure}[ht]
\includegraphics[scale=0.45,angle=0]{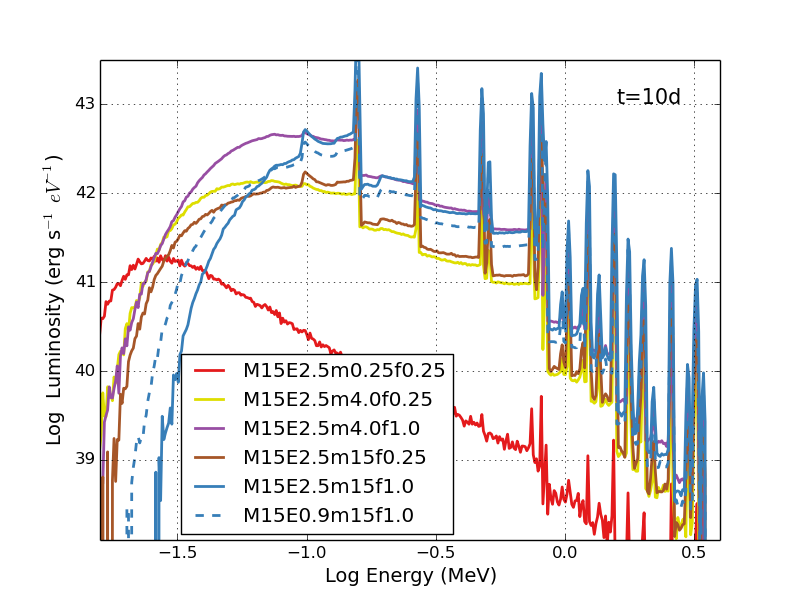}
\includegraphics[scale=0.45,angle=0]{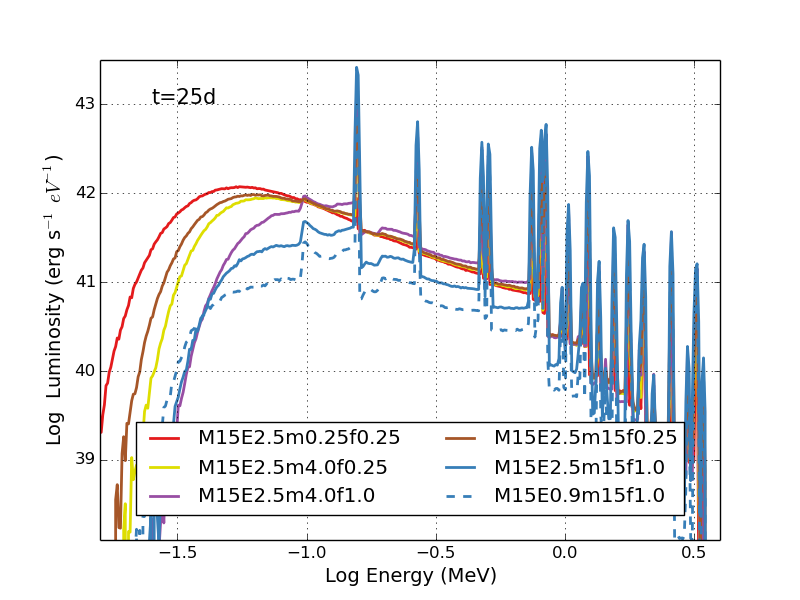}
\includegraphics[scale=0.45,angle=0]{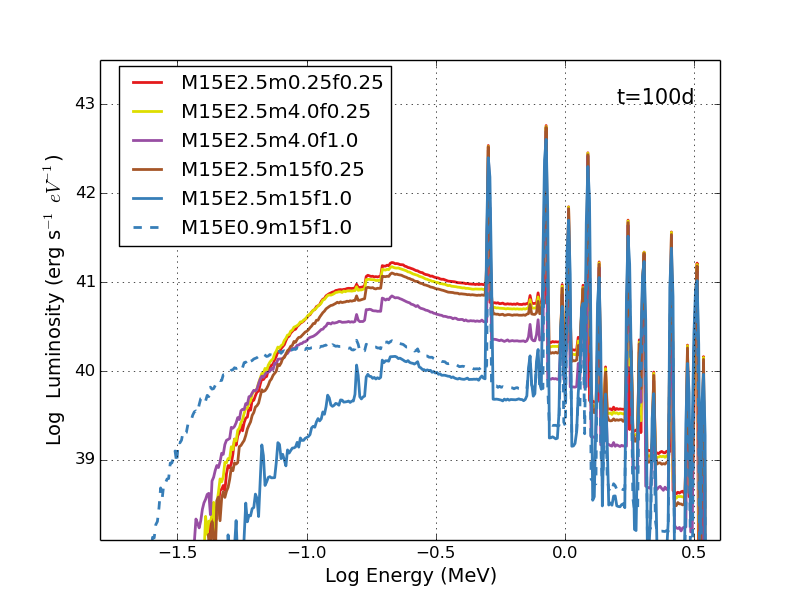}
\includegraphics[scale=0.45,angle=0]{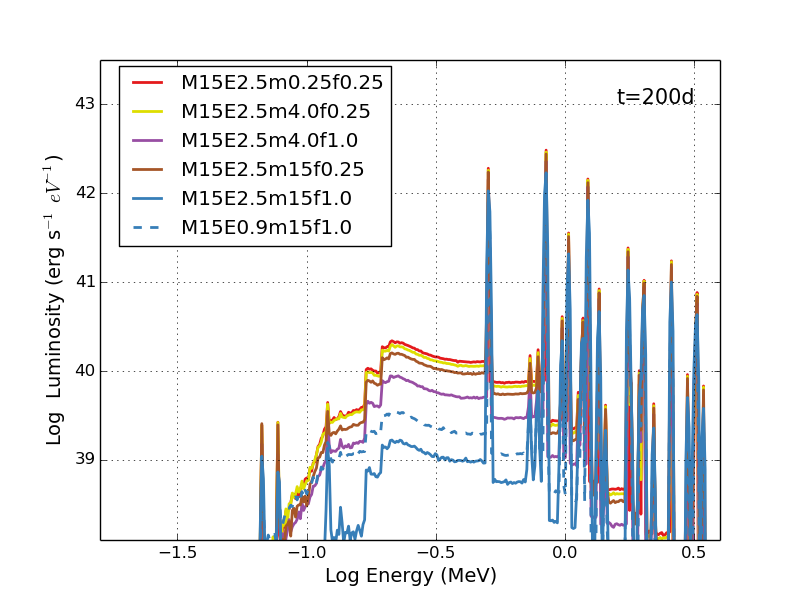}
\caption{X-ray and gamma-ray spectra for 5 of our mixed explosion models:  4 models of our strong 15\,M$_\odot$ explosion with different mixing properties and one weak explosion.  The shape of the continuum and its evolution allows us to differentiate the different models.} 
\label{fig:grayspectra}
\end{figure}

Figure~\ref{fig:grayspectra} shows the spectra at four different times from five of these models.  Our 20 and 25\,M$_\odot$ cases did not eject much $^{56}$Ni and have very weak signals and we focus the study here on the 15\,M$_\odot$ explosions. Even with the same $^{56}$Ni yield, the continuum and line flux varies by a factor of 10.  For the models with extended mixing, the line features at early times will be strong.  The models with low mixing only produce such strong lines at late times when the $^{56}$Ni in their explosions can escape.  The low-energy continuum emission, arising for many scatterings of the MeV photons, quickly becomes strongest for the models with little mixing (where the gamma-rays must scatter, losing energy, many times before escaping).

The efficacy of individual spectra to differentiate models depends on observation epoch and resolution of the acquired spectra (i.e., Figure~\ref{fig:grayspectra}). Line intensities and their profiles can be characterized using high-resolution gamma-ray spectroscopy; sensitivity limitations require a Galactic supernova. In addition, detailed line-based population studies are precluded by the low SN event rate, at least for the current generation of space-based telescopes (e.g., \citet{2019BAAS...51g..98T}).

\begin{figure}[ht]
\includegraphics[scale=0.45,angle=0]{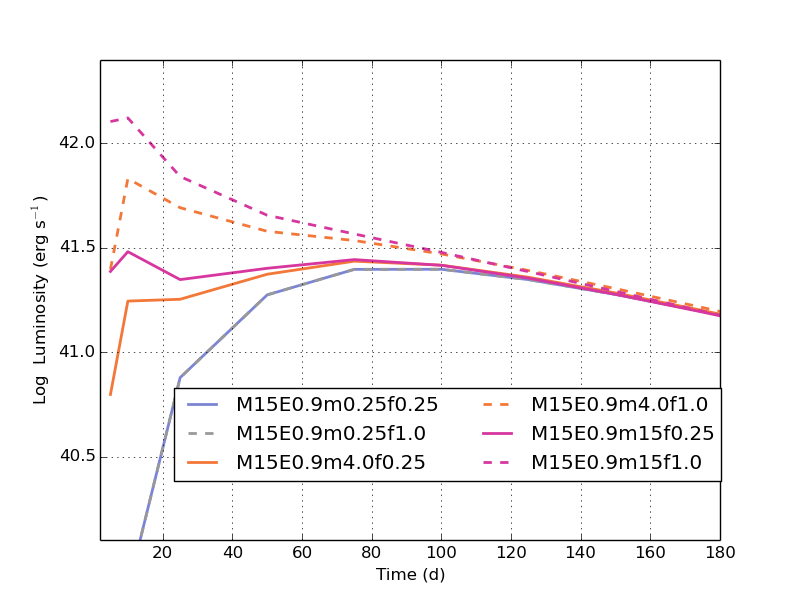}
\includegraphics[scale=0.45,angle=0]{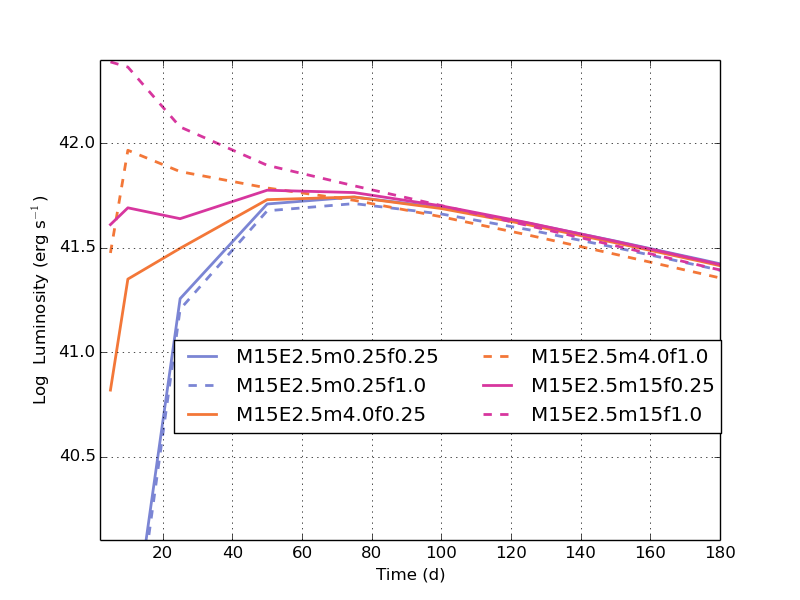}
\caption{Gamma-ray luminosity (photons above 100keV) as a function of time for our suite of 15\,M$_\odot$ explosions.  The gamma-ray lightcurve is sensitive to both the extent and amount of mixing.  The rise time depends on the extent and the luminosity depends more sensitively on the amount of mixing.  Late time signals are sensitive to the total $^{56}$Ni yield.} 
\label{fig:graylc}
\end{figure}

Time evolution is a valuable alternative diagnostic tool. Figure~\ref{fig:graylc} shows the gamma-ray luminosity (photon energies above 100\,keV) as a function of time for our different 15\,M$_\odot$ explosions. The interaction of nuclear line photons with the ejecta leads to a gamma-ray continuum that is present at all epochs. Here, all electrons serve as scattering targets at the energies of interest, independent of the thermal environment or ionization states that affect emergent spectra in the X-ray and UV-optical-infrared (UVOIR) bands. The magnitude and shape of the continuum, as well as the nuclear lines, are time-dependent.

The early-time emission, particularly in the first 20--40 days, is sensitive to the extent and amount of the mixing. With early observations the extent of the mixing can be determined from the light curve rise time, or the early-time luminosity if the source distance is known. At late epochs gamma rays can escape freely as the ejecta becomes optically thin; in this late regime the light curves are dominated by nuclear lines and the intensity reflects the amount of $^{56}$Co (and hence $^{56}$Ni) produced. Thus, light-curves spanning extended post-explosion epochs are excellent diagnostic probes of mixing and elemental abundance yields; monitoring the spectral and temporal evolution of gamma-ray emission (both lines and continuum), from early through late post-explosion epochs, contributes to a complete picture that can facilitate model differentiation (e.g., Miller et al. 2023, in preparation).

\section{Summary and Discussion on the Diagnostic Synergies}
\label{sec:corr}

Core-collapse supernova explosions play a wide role in astrophysics by producing compact remnants (neutron stars, black holes) and the synthesis and injection of many heavy elements into the Galaxy.  Because they are produced in some of the most extreme conditions in the universe, they can also probe the physics in extreme conditions (matter at nuclear densities, neutrinos, etc.).  Although the evidence that the convection-enhanced supernova engine~\citep{herant94} is the primary engine behind these explosions, alternative mechanisms have been proposed and, at some level, probably occur in Nature.  Understanding these engines not only allows astronomers to both determine the properties of supernovae and their products, but also better use these cosmic explosions to probe the extreme physics.  This understanding is gained both through concurrent measurements where multiple diagnostics observe a single event as well as by combining information from a broad set of events.    In this section, we discuss both how individual diagnostics complement each other and the interplay between concurrent observations of a single event and single-diagnostic observations of a large set of events.  The diagnostics and their synergies are summarized in table~\ref{tab:synergies}.  For concurrent observations, the timescales are included to provide an indication of the coordination and preparedness needed to obtain joint observations.

\begin{table}[ht]
    \begin{center}
    \begin{tabular}{lccc}
    \hline
    Diagnostic & Milky Way & Andromeda & Population \\
    $t_{\rm peak}$-$t_{\rm bounce}$ & & & \\
    \hline
    Gravitational Waves (GW)& Convection & Rotation & \\
   0 & & & \\
    Neutrinos ($\nu$) & Bounce properties & Bounce? & \\
    0 & & & \\
    Gamma-Rays ($\gamma$-ray) & Exp. Asym. & Exp. Asym. & \\
    5-100\,d & Calibrated by GW and $\nu$ & Compared to Shock Breakout & \\
    \hline
    Shock Breakout (SBO) & Exp. Asym. & Exp. Asym. & Exp. Asym. (within $\sim 200\, {\rm Mpc}$) \\
     30-300\,s (Ib/c) & Calibrated by GW and $\nu$ & Compare to $\gamma$-rays & compare to LC, CNR, CR \\
     3000-10,000\,s (IIp) & & & \\
    UVOIR (LC) & Exp. Asym. (late-time obs.)? & Mixing (late-time obs)? & Exp. Asym. (metallicity evolution)? \\
     10-20\,d & Calibrated by GW and $\nu$ & Calibrated by $\gamma$-rays and SBO & compare to SNR, SBO, CR \\
     late-time obs ($>100$\,d) & & & \\
     \hline
     Supernova Remnants (SNR) & & & Exp. Asym. for MW SNR \\
     & & & compare to SBO, LC, CR \\
     Compact Remnants (CR) & & & convection (metallicity evolution) \\
     & & & compare to SBO, LC, SNR \\
    \hline
    \end{tabular}
    \caption{For each diagnostic, the capability of the probe is described and if it is calibrated by other observations.   Some diagnostics probe the convective engine or properties of the bounce (e.g. neutron star properties) directly.  Many probe explosion asymmetries that, in turn, can be used to probe the supernova engine.  If blank, the probe is not effective for the conditions suggested.  For concurrent diagnostics, we provide the rough times post-bounce of peak emission.  This gives an idea of the coordination and preparation needed for joint observations.}
    \end{center}
    \label{tab:synergies}
\end{table}

\subsection{Concurrent Observations:}

Concurrent observations are ideally suited to testing the efficacy of and calibrating diagnostics that then can used in broader studies where only a single diagnostic is measured.  However, as we discussed in this paper, many of the most direct diagnostics are limited to nearby events.  Here we give two examples (based on the distance of the supernova) of how a well-studied event can calibrate the use of far-reaching diagnostics to study supernova populations (next section). \\

\noindent
{\it Galactic Event (within $\sim 10\,{\rm kpc}$)} --- A Galactic event can leverage the full power of our direct diagnostics.  By comparing the detailed direct observations to features in our indirect, but more far-reaching diagnostics, we will be better able to use these indirect observations to study supernova.  Let's review these diagnostics with these connections in mind.

Especially with next generation detectors, gravitational wave observations can measure the growth time and magnitude of the convection above the proto-neutron star.  This physics is at the heart of our current understanding of the supernova.  At this distance, gravitational waves will be able measure even modest rotation in the collapsing core, determining the role of rotation in the explosion mechanism.  Comparing these direct observations with other diagnostics can strengthen the interpretation of all supernova events:
\begin{itemize}
    \item {\it Neutrinos:}  In addition to probes of nuclear physics (e.g. neutrino oscillations), neutrino observations of a Galactic event will probe the bounce structure.  These observations will constrain the supernova models, and hence the interpretation of the gravitational wave observations of both rotation and convection.
    \item {\it Prompt Gamma-ray Observations:}  Gamma-ray observations of a Galactic supernova can be used to both probe the structure of the star (using a variety of radioactive isotopes) and the mixing of radioactive Ni, and hence, asymmetries in the supernova explosion~\citep{2020ApJ...890...35A}.  Compared to gravitational wave observations, this mixing can be tied directly to the supernova engine.
    \item {\it Shock Breakout:}  The shock breakout signal is driven by a broad set of effects (asymmetries in the explosion, stellar structure, stellar wind).  These effects can be disentangled in our shock breakout models by comparing shock breakout observations of a nearby event to both gravitational waves and gamma rays.
    \item {\it Supernova Lightcurves and Spectra:}  Galactic supernovae can be studied in detail and, even though these observations are not extremely sensitive to the details of the explosion, the fidelity of the observations, coupled with comparisons to other diagnostics will help tie features in this diagnostic to the supernova engine.  As we have shown, e.g. in Figure~\ref{fig:lc15}, the evolution of the light curves at late-times (falling off of the plateau phase) can be used to probe the mixing.  The spectra vary as well at these late times and can be used to distinguish mixing.  Much more work is needed to improve our models to fully identify spectral features that probe mixing.  At the rate of such Galactic events and the model uncertainties in this emission, it may be difficult to truly calibrate this diagnostic from Galactic events.
    \item {\it Supernova Remnants:}  Observations of supernova 1987A have proven how difficult it is to determine the fate (neutron star or black hole) of the compact remnant~\citep{1999ApJ...511..885F,2018ApJ...864..174A,2020ApJ...898..125P}.  With better neutrino detectors, the fate of the remnant can be determined through neutrinos~\citep{2021PhRvD.103b3016L}, placing some constraints on the engine~\citep{2021PhRvD.103b3016L}.  But getting compact remnant masses which place more stringent constraints on the engine will require a different set of observations.
\end{itemize}
The strength of coupling these observations lies in using the observations to refine and improve models of the diagnostics.   The efficacy of such comparisons that tie these diagnostics together relies on a close interaction with a broad range of supernova models.

{\it Supernovae out to the Andromeda Galaxy ($\sim 1\,{\rm Mpc}$)}

If we move to distances comparable to the Andromeda galaxy, we increase the number of detected supernovae, but diminish the power of our direct diagnostics due to reduced signal to noise in the detection.  Gravitational waves can be used to rule out highly-rotating events, providing limits on the number of rotationally-powered supernovae.  This could not be studied easily with a small number of Galactic events because, if the standard theory (both stellar evolution and supernova engine models) is correct, such rotationally-powered supernovae should be rare (10-20\% of all events at most).  Shock breakout and gamma-ray measurements are easily doable at 1\,Mpc distances if next-generation detectors are developed.  These measurements can both be used to calibrate supernova light-curves and spectra measurements, but also develop a better understanding of supernova populations to understand extended observations.

The current and funded suite of gamma-ray telescopes, including Fermi~\citep{2009ApJ...697.1071A}, Integral~\citep{2003A&A...411L.131U} and COSI~\citep{2019BAAS...51g..98T} are all capable of these transient observations and will detect nearby supernova events, but not out to this distance.  A number of mission concepts have been proposed to increase the distance of detection of radioactive decay emission from supernovae:  AMEGO-X~\citep{2022JATIS...8d4003C}, LOX~\citep{2019LPICo2135.5055M}, (Miller et al., submitted), and ASCENT~\citep{2023arXiv230101525K}.  Similarly, detectors such as {\it UltraSAT}~\citep{2022arXiv220800159B} and CASTOR~\citep{2017XRS....46..303M} satellites are designed to increase the number of shock breakout detections in the UV.  Whether these missions will provide sufficient information to help us understand the supernova engine remains to be determined, but it is more likely that to disentangle all of this physics, joint UV and X-ray observations are required such as proposed by the {\it SIBEX} mission~\citep{2018FrASS...5...25R}.  In both cases considerable modeling work is needed if we are to use these observations to use these supernova observations to both constrain the engine and make predictions for extended populations.

\subsection{Single Diagnostic Observations of Supernova Populations}

Many diagnostics are not measured concurrently with other diagnostics.  These diagnostics include compact remnant properties, ejecta remnants, galactic chemical evolution, supernova light-curves and spectra.  Although the diagnostics are not concurrent, they still constitute a multi-messenger approach.  Although a single, well-studied Galactic event can provide a huge amount of insight into the supernova engine, single events can be abnormal (SN 1987A, for example, was not a normal supernova explosion).  To truly understand the supernova engine, diagnostics that study broad populations of events are critical.  Such studies often require refined theory models and these theoretical models must work both to explain nearby, well-studied events and single-diagnostic populations.  In this final section, we describe a few examples of how these diagnostics of supernova populations.

Supernova light-curves and spectra are one of the few diagnostics of large populations that are also constrained by nearby events.  The models of these systems can be tested against nearby events and then applied to the increasingly growing sample of observations.  Because these observations can be calibrated against nearby observations, an empirical approach can be used (without even invoking model simulations of this diagnostic) to constrain the supernova engine.  But the constraints will prove stronger if a strong theoretical understanding can be coupled to the observed populations.

Supernova ejecta remnants can also be weakly calibrated to nearby observations.  Supernova light-curves and spectra as well as gamma-ray emission can all provide some insight into the yields and yield distributions from supernova explosions.  These observations can be used to study supernova ejecta remnants empirically, but it is likely that these studies will continue to use theoretical models, calibrated by observations, to tie ejecta remnant observations to the explosion mechanism.

Although some information of compact remnants can be obtained through nearby events, data on spins and mass distributions (which can be tied to the supernova engine through theoretical models) arise more through large population studies (e.g. gravitational wave measurements).  For this diagnostic, the tie between other diagnostics is almost entirely through theoretical models.

In all three of these examples (light-curves, ejecta remnants and compact remnants), theoretical models strengthen the synergy between diagnostics.  These diagnostics all contribute to the multi-messenger constraints of the supernova engine.

\begin{acknowledgements}

The work by CLF, ALH was supported by the US Department of Energy through the Los Alamos National Laboratory. Los Alamos National Laboratory is operated by Triad National Security, LLC, for the National Nuclear Security Administration of U.S.\ Department of Energy (Contract No.\ 89233218CNA000001).  A portion of the work by CLF was performed at the Aspen Center for Physics, which is supported by National Science Foundation grant PHY-1607611.

\end{acknowledgements}

\bibliography{refs}{}
\bibliographystyle{aasjournal}

\end{document}